\documentclass[10pt,journal,compsoc]{IEEEtran}
\ifCLASSOPTIONcompsoc
  \usepackage[nocompress]{cite}
\else
  \usepackage{cite}
\fi
\ifCLASSINFOpdf
\else
\fi
\usepackage{cite}
\usepackage{dutchcal}
\usepackage[T1]{fontenc}
\usepackage{subfigure}
\usepackage{textcomp}
\usepackage{xcolor}
\usepackage{booktabs}
\usepackage{footnote}
\usepackage{threeparttable}
\usepackage{algorithm}
\usepackage{algorithmic}
\usepackage{multirow}
\usepackage{graphicx}
\makeatletter

\newcommand{\Rmnum}[1]{\expandafter\@slowromancap\romannumeral #1@}
\usepackage{amsmath,amssymb,amsfonts}
\makeatother
\begin{document}
%
\title{Null Model-Based Data Augmentation \\ for Graph Classification}
%
%
%
%

\author{Qi~Xuan,~\IEEEmembership{Senior Member,~IEEE},~Zeyu~Wang,~Jinhuan~Wang,~Yalu~Shan,\\
Xiaoke Xu,~\IEEEmembership{Member,~IEEE},~Guanrong Chen,~\IEEEmembership{Fellow,~IEEE}
\IEEEcompsocitemizethanks{
\IEEEcompsocthanksitem Q. Xuan, Z. Wang, J. Wang, Y. Shan are with the Institute of Cyberspace Security, 
College of Information Engineering, Zhejiang University of Technology, Hangzhou, China (E-mail: xuanqi@zjut.edu.cn; Vencent\_Wang@outlook.com; jhwang@zjut.edu.cn; yalushan@foxmail.com).\protect
\IEEEcompsocthanksitem X. Xu is with the College of Information and Communication Engineering, Dalian Minzu University, Dalian, China (E-mail: xuxiaoke@foxmail.com).\protect
\IEEEcompsocthanksitem G. Chen is with the Department of Electrical Engineering, City University of Hong Kong, Hong Kong SAR, China (E-mail: eegchen@cityu.edu.hk).\protect
\IEEEcompsocthanksitem Corresponding author: Jinhuan Wang.
}
}
\IEEEtitleabstractindextext{%
\begin{abstract}
In network science, the null model is typically used to generate a series of graphs based on randomization as a term of comparison to verify whether a network in question displays some non-trivial features such as community structure. Since such non-trivial features play a significant role in graph classification, the null model could be useful for network data augmentation to enhance classification performance. In this paper, we propose a novel technique that combines the null model with data augmentation for graph classification. 
Moreover, we propose four standard null model-based augmentation methods and four approximate null model-based augmentation methods to verify and improve the performance of our graph classification technique.
Our experiments demonstrate that the proposed augmentation technique has significantly achieved general improvement on the tested datasets. In addition, we find that the standard null model-based augmentation methods always outperform the approximate ones, depending on the design mechanisms of the null models. Our results indicate that the choice of non-trivial features is significant for increasing the performance of augmentation models for different network structures, which also provides a new perspective of data augmentation for studying various graph classification methods.
\end{abstract}

}

\maketitle

\IEEEdisplaynontitleabstractindextext

%
\IEEEpeerreviewmaketitle

\IEEEraisesectionheading{\section{Introduction}\label{sec:introduction}}

%
%
%
%

 

\IEEEPARstart{N}{ull} models are pattern-generating models that deliberately exclude a mechanism being tested and primarily rely on its ability to explore non-trivial features of graphs, as a popular analytical tool applied to investigating the dynamics of complex networks. 
Null models has been applied to analyzing ecological and biogeographic data and quantifying complex network properties such as community structure~\cite{newman2006finding, cazabet2017enhancing}, assortativity~\cite{pastor2001dynamical}, degree correlation~\cite{mahadevan2006systematic}, epidemic spreading rate~\cite{estrada2016epidemic}, routing efficiency~\cite{nian2014efficient}, pattern detection~\cite{ulrich2013pattern}, microbial diversification~\cite{zhai2018null}, etc. 
In many applications, null models can reveal important network properties that could not be directly quantified by other models or methods. Inspired by this, we employ null models to improve the accuracy of graph classification.

\begin{figure}[!t]
  \begin{center}
  \includegraphics[width=.9\linewidth]{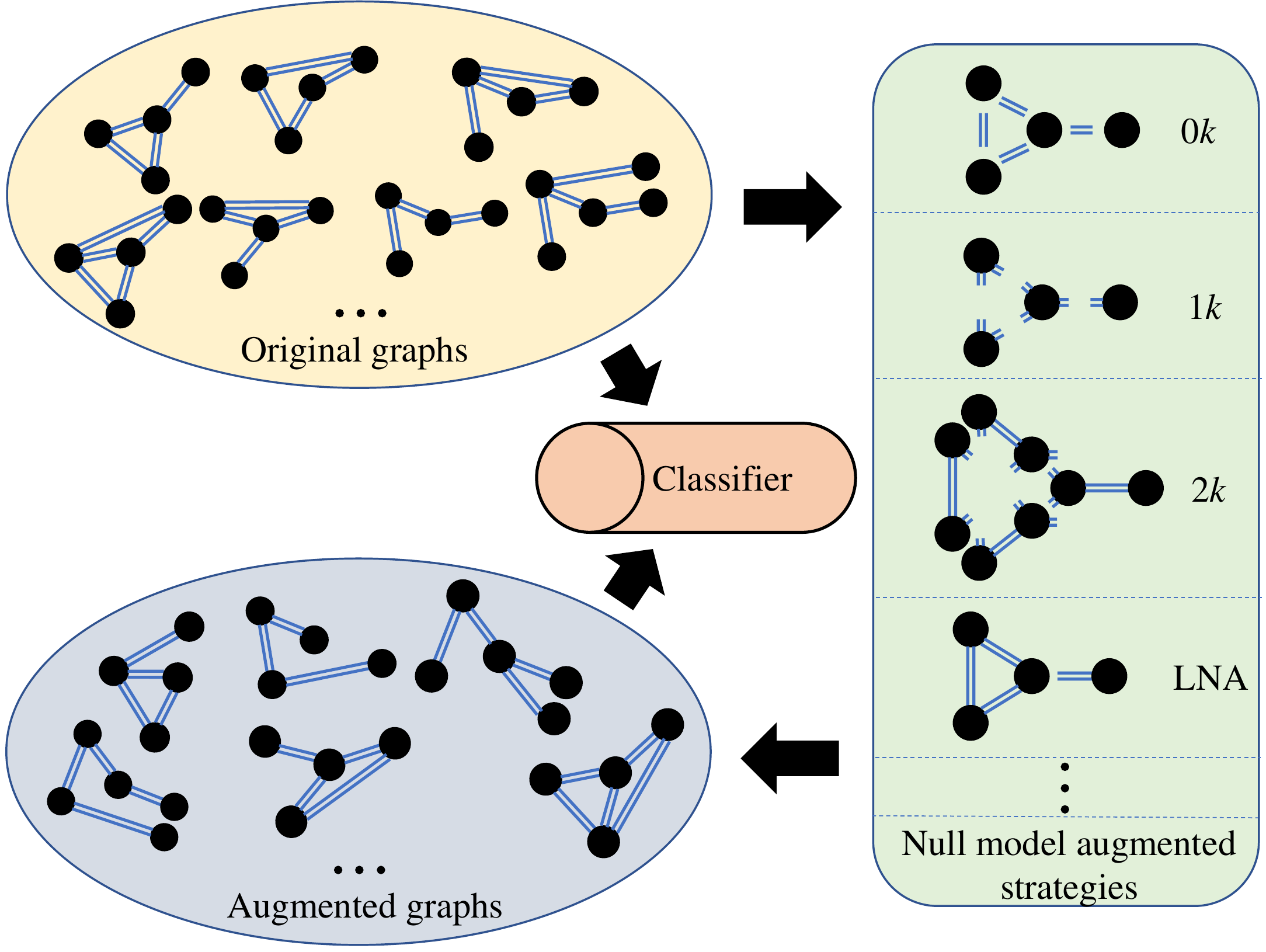}
  \caption{The process of data augmentation based on null model. }
  \label{fig:augmentation}
  \end{center}
\end{figure}
Various, the graph classification augmentation methods can be roughly divided into two categories: feature augmentation methods and data augmentation methods. Feature augmentation methods focus on the embedding obtained from classical graph classification methods. Specifically, this kind of methods utilize feature selection and feature splicing. In \cite{2010FeatureSelection}, a semi-supervised feature selection approach is taken to search for optimal subgraph features with labeled and unlabeled graphs so as to improve graph classification performance. In \cite{xuan2021subgraph}, the concept of subgraph network (SGN) is introduced and used to expand the structural feature space of the underlying network, thereby enhancing network classification. In \cite{wang2021sampling}, a sampling subgraph network is constructed to address the problem of the SGN model that lacks diversity and has high time complexity. Whereas a feature fusion framework is also established to integrate the structural features of diverse sampling subgraphs to improve the performance of graph classification. 

Data augmentation, on the other hand, is a technology that artificially expands the training datasets by allowing limited data to generate more equivalent data, so as to improve the performance of downstream tasks, such as node classification\cite{zhao2020data}, community detection\cite{ZhouRobustECD}, and graph classification\cite{zhou2020m}. Researches in network science, especially in graph classification, focused on the graph structure and proposed some heuristic methods from the perspective of nodes or communities to modify the graph topology structure~\cite{zhou2020data,kong2021flag}. These methods generate augmented data by introducing a tiny disturbance into the original data and then altering the parameters through retraining the model, which can also be thought of as a regularization method but lacks the instruction of data augmentation. 
Graph classification aims to identify the category labels of graphs in a dataset by using features that are extracted by handcraft, graph kernels~\cite{siglidis2020grakel}, graph embedding~\cite{goyal2018graph}, or graph neural networks~\cite{scarselli2008graph}. Some non-trivial features of the graph are identified by the classifier. Practically, one should pay more attention to these non-trivial features when proposing data augmentation methods, which not only improve the classifier performance, but also provide some inspiration for the interpretablity of the graph classification.

In this paper, we propose a new technique that combines the null model with data augmentation for graph classification. The idea is to find non-trivial features of the graph and develop a null model according to the non-trivial features to produce more virtual data for retraining the graph classifier, as illustrated in Fig.~\ref{fig:augmentation}. Since the virtual data we generated guarantees that the non-trivial features will not change, their availability remains to be verified. Given this, we adopt the data filter proposed in \cite{zhou2020m} to filter out fine augmented examples from the generated data. We demonstrate that the method can significantly improve the performance of graph classification. Specifically, we have the following contributions:
\begin{itemize}
    \item We propose a new technique for graph data augmentation. It is the first methodology to combine the null model with data augmentation and applied to graph classification tasks. 
    

    
    \item We develop four standard null model-based data augmentation models and four approximate null model-based data augmentation methods for non-trivial features, which play a significant role in the graph classification.
    
    \item We apply our data augmentation methods based on several graph classification methods, and our experimental results on five real-world network datasets demonstrate the effectiveness of the proposed methods.  
\end{itemize}

The rest of the paper is organized as follows. In Sec.~\ref{sec:related}, we present a brief description of the null model, data augmentation, and graph classification methods. In Sec.~\ref{sec:method}, we introduce four standard null model-based data augmentation methods 
and four approximate null model-based data augmentation methods. In Sec.~\ref{sec:expset}, we describe our experimental settings and discuss the results in detail. Finally, in Sec.~\ref{sec:conclusion}, we conclude the paper and outline some future work. 

\section{Related Work}\label{sec:related}
In this section, we give a brief review on related work of null models, data augmentation, and graph classification algorithms in graph data mining\cite{xuan2021graphdatamining}.

\subsection{Null Model}


A null model is a pattern-generating model based on random sampling from a known or imagined distribution~\cite{gotelli1996null}. Null model as an analysis tool was applied in ecological and biogeographic studies in the past few decades, such as the laws of species migration, island rules, and the spatial patterns of trees in temperate forests~\cite{biddick2021simple,carrer2018tree,reimann2019null}. Different from other generating models, a null model constructs a model that deliberately excludes a mechanism being tested~\cite{gotelli2001research}. Many null models are used to simulate different processes and find the underlying rules, demonstrated that the null models can explain the internal mechanism of the tested model. In network analysis, a null model consists of a network that can be seen as one specific graph with some structural attributes but otherwise is treated as a random network instance. For undirected graphs, several null model generation methods have been proposed, such as 0-order, 1-order, 2-order, 2.5-order null model graphs~\cite{gjoka20132,mahadevan2007orbis,mahadevan2006systematic,newman2001random,chung2002connected,chung2002average}. Such null models are helpful to explore the nature of modeling and structures of complex networks. However, it is noticed that there is no study to analyze the graph classification task by utilizing null models. In this paper, we adopt three null models with different orders and five novel (approximate) null models generated according to different non-trivial features to enhance the performance of graph classification.

\subsection{Data Augmentation}
Data augmentation is a common method for solving problems caused by insufficient dataset or model overfitting, and is widely applied to computer vision~\cite{cubuk2019autoaugment,devries2017improved} and natural language processing~\cite{fadaee2017data,csahin2019data}. Nevertheless, data augmentation in graph data is still in its infancy. In~\cite{zhou2020m}, three heuristic methods are developed to generate the virtual data and achieve average improvement in accuracy on graph classification tasks. For node classification tasks, in~\cite{zhao2020data}, it was discovered that neural edge predictors can effectively encode class-homophilic structure to promote intra-class edges and demote inter-class edges in given graph structure and they leveraged these results to improve the performance of GNN-based node classification via edge prediction. In~\cite{dong2020data}, the labels of the training set are propagated through the graph structure expanding the training set. Although the above methods can improve the classification performance by expanding the dataset, they all lack interpretability. In this paper, we propose a new technique that combines the null model framework with data augmentation, which can provide an explanatory basis for data augmentation in graph classification. 

\subsection{Graph Classification}
Graph data is ubiquitous in nature and human society, ranging from atomic structures to social networks. Graph classification is an important task in the graph data mining, where its objective is to predict the labels of networks correctly. This task usually is implemented by combining machine learning classifiers and graph representation learning such as graph kernel, graph embedding, and deep learning methods. Among them, the graph kernel is a graph representation method directly oriented to the graph structure, which contains structured information of higher-dimensional Hilbert space, such as WL kernel~\cite{2011Weisfeiler} and Deep WL~\cite{2015Deep}. While graph embedding is a method that maps networks
to low micro-dense vectors, such as graph2vec~\cite{narayanan2017graph2vec}. Deep learning methods have attracted much attention in recent years, including graph neural network (GNN)~\cite{GNN}, graph convolutional network (GCN)~\cite{GCN}, and Diffpool~\cite{2018Hierarchical}. Although the above graph representation methods have relatively high expressiveness and learning ability, they don't have good interpretability. Moreover, they rely only on a single network structure, limiting their ability to exploit the latent structural features. In this paper, we combine the interpretability of null models with high expressiveness graph representation methods. Our experimental results demonstrate that the proposed null models indeed enhance the graph representation methods.

\section{Methodology}\label{sec:method}

In this section, we first formulate the problem of data augmentation on graph classification. We then summarize ten graph attributes, propose four standard null model-based augmentation (0$k$, 1$k$, 2$k$, LNA) and four approximate null model-based augmentation methods ADA-(C, \textsc{Bc}, \textsc{Cc}, \textsc{Ec}), and demonstrate the construction of the algorithms.

\begin{table}[!t]
  \centering
  \caption{Notations used in this paper.}
  \resizebox{1\columnwidth}{!}{
    \renewcommand\arraystretch{1.1}
    \begin{tabular}{lr}
    \toprule
    Symbol & Definition \\
    \midrule
    $G,G_{aug}$ & Original/augmented graph \\
    $V,E$ & Set of nodes/edges in graph $G$ \\
    $n,m$   & Number of nodes/edges \\
    $\mathcal{D}$ & Average  degree of graph $G$ \\
    $C$ & Average clustering coefficient of graph $G$ \\
    $P_L$ & The proportion of leaf nodes of graph $G$ \\
    $d_{max}$ & The maximun degree of graph $G$ \\
    $P_D, J_D$ & The (joint) degree distribution of graph $G$ \\
    $E_C$ & Average eigenvector centrality of graph $G$ \\
    $B_C$ & Average betweenness centrality of graph $G$ \\
    $C_C$ & Average closeness centrality of graph $G$ \\
    $E_{leaf}$ & The set of edges with leaf nodes \\
    $E'$ & The edges set of augmented graph $G'$ \\
    $e_{add}, e_{del}$ & Edge of addition/deletion \\
    $E_{add}, E_{del}$ & Edges set of addition/deletion \\ 
    $\mathcal{S}$ & The set of node feature value \\
    $f_i$ & The feature($C$/$B_C$/$C_C$/$E_C$) value of node $v_i$ \\
    $F$ & The feature($C$/$B_C$/$C_C$/$E_C$) value of graph $G$ \\
    $\alpha$ & The cost coefficient of augmentation\\
    $T$ & Approximate augmentation iterations\\
    $\mathcal{F}$ & The augmentation function\\
    \bottomrule
    \end{tabular}%
  }
  \label{tab:nations}%
\end{table}%

\subsection{Notation and Problem Statement}

Let $D$=$\{(G_i,y_i)|i=1,2,\cdots,N\}$ denote a graph dataset that has $N$ undirected unweighted graphs, where $G_i$ is a graph with label $y_i$. As usual a graph is denoted as $G=(V,E)$ where 
$V$=$\{v_j|j=1,2,\cdots,n\}$ and $E$=$\{e_j|j=1,2,\cdots,m\}$ are the node and edge sets, with $n$ and $m$ denoting the numbers of nodes and edges, respectively. Specifically, for a dataset $D$, it is split to training set $D_{train}$, validation set $D_{val}$ and testing set $D_{test}$. Among them, the training set $D_{train}$ and validation set $D_{val}$ are used to pre-train the classifier and yield the original graph classifier $\mathcal{C}_{ori}$.

%

Graph data augmentation aims to generate novel and realistically rational graphs by designing a transformation model $G_{aug}$ $\sim$ $\mathcal{F}(G_{aug}|G)$, where $\mathcal{F}(\cdot|G)$ is the augmentation distribution conditioned on the original graph, representing the prior information for data distribution. Here, we adopt null models for constructing augmented graphs. Through data augmentation for each graph $G$ in the training set, we get the augmented set $D_{aug}$=$\{G_{aug}\}$. 
Then, we combine $D_{train}$ and $D_{aug}$ to get $D'_{train}$,

\begin{equation}
  D'_{train} = D_{train} + D_{aug}, \label{eq:Augmented Set}
\end{equation}
which will be used to retrain the classifier with optimization, and so as to obtain the augmented classifier $\mathcal{C}_{aug}$. 

\subsection{Graph Attributes}
Graph attributes reflect the topology of a graph from different perspectives, which contribute to the graph classification. Here, we summarize ten classical graph attributes.

\begin{itemize}
  \item \textbf{Number of nodes ($n$)}: The number of nodes in the graph $G$.
  \item \textbf{Number of edges ($m$)}: The number of edges in the graph $G$.
  \item \textbf{Average degree ($\mathcal{D}$)}: The degree of a node in a graph is the number of connections it has to other nodes. The average degree of the graph $G$ is given by
  \begin{equation}
    \mathcal{D} = \frac{2m}{n}.
    \label{eq:Average degree of graph}
  \end{equation}
  
  \item \textbf{Degree distribution ($P_D$)}: The degree distribution is the probability distribution of all degrees over the whole network. The degree distribution of the graph $G$ is calculated by
  \begin{equation}
    P_D(k) = \frac{n_k}{n},
    \label{eq:Average degree distribution of graph}
  \end{equation}
  where $n_k$ denotes the number of nodes with degree $k$ in $G$.
  \item \textbf{The proportion of leaf nodes ($P_L$)}: Leaf node is the node with degree one. The proportion of leaf nodes in the graph $G$ is calculated by
  \begin{equation}
    P_L = \frac{n_L}{n},
    \label{eq:proportion of leaf nodes of graph}
  \end{equation}
  where $n_L$ denotes the number of leaf nodes in $G$.
  

  \item \textbf{Joint degree distribution ($J_D$)}: The joint degree distribution~\cite{joint_degree_distribution} refers to the number of degree (probability) of the nodes of each edge. Here, the joint degree distribution of the graph $G$ is defined as
  \begin{equation}
    J_D(k_1,k_2) = \frac{\mu(k_1,k_2)m(k_1,k_2)}{2m}, 
    \label{eq:joint degree distribution}
  \end{equation}
  where $m(k_1,k_2)$ denotes the number of edges that the nodes with degree $k_1$ and the nodes with degree $k_2$, and $\mu(k_1,k_2)$ is defined as
  \begin{equation}
  \mu(k_1,k_2) =
  \begin{cases} 
  2, &k_1=k_2 \\
  1, &otherwise
  \end{cases}
  \label{eq:coefficient of joint degree distribution}
  \end{equation}

  \item \textbf{Average clustering coefficient ($C$)}: The clustering coefficient~\cite{borgatti1997network, lind2005cycles} is a coefficient used to describe the degree of clustering between nodes in the graph. The average clustering coefficient of the graph $G$ can be denoted as
  \begin{equation}
    C = \frac{1}{n}\sum_{i=1}^n{\frac{2L_i}{k_i(k_i-1)}},
    \label{eq:clutering}
  \end{equation} 
  where the $k_i$ is the degree of node $v_i$ and $L_i$ is the number of edges among the $k_i$ neighbors of node $v_i$.
  \item \textbf{Average betweenness centrality ($B_C$)}: The betweenness centrality~\cite{wasserman1994social} is a measure of graph centrality based on the shortest paths. The average betweenness centrality of the graph $G$ is coomputed by
  \begin{equation}
    B_C = \frac{1}{n}\sum_{i=1}^{n}{\sum_{s\neq i\neq t}{\frac{m_{st}^i}{g_{st}}}},
    \label{eq:betweenness centrality}
  \end{equation}
  where $g_{st}$ is the number of shortest paths between node $v_s$ and node $v_t$, and $m_{st}^i$ is the number of shortest paths passing through node $v_i$ between node $v_s$ and node $v_t$.
  \item \textbf{Average closeness centrality ($C_C$)}: The closeness centrality~\cite{bavelas1950communication,beauchamp1965improved} is the average length of the shortest paths between a node and other nodes in the graph. And the average closeness centrality of the graph $G$ is computed by
  \begin{equation}
    C_C = \frac{1}{n}\sum_{i=1}^{n}{\frac{n}{\sum_{j=1}^{n}{d_{ij}}}},
    \label{eq:closeness centrality}
  \end{equation}
  where $d_{ij}$ is the lehgth of shortest path between nodes $v_i$ and $v_j$.
  \item \textbf{Average eigenvector centrality ($E_C$)}: The eigenvector centrality~\cite{bonacich1972factoring,bonacich2007some} is used to measure the importance of nodes on the network. The average eigenvector centrality of the graph $G$ is defined as
  \begin{equation}
    E_C = \frac{1}{n}\sum_{i=1}^{n}{x_i},
    \label{eq:eigenvector centrality}
  \end{equation}
  where $x_i$ is the importance score of node $v_i$ and given by:
  \begin{equation}
      x_i = \lambda \sum_{j=1}^{n}{A_{ij}x_j}, 
      \label{eq:importance measure x}
  \end{equation}
  where $\lambda$ is an adjustable parameter, which should be less than the reciprocal of the maximum eigenvalue of the adjacency matrix $A$. Whereas $A_{ij}$ represents the weight of the edge between node $v_i$ and node $v_j$. 
\end{itemize}

\begin{figure}[!t]
  \begin{center}
  \includegraphics[width=\linewidth]{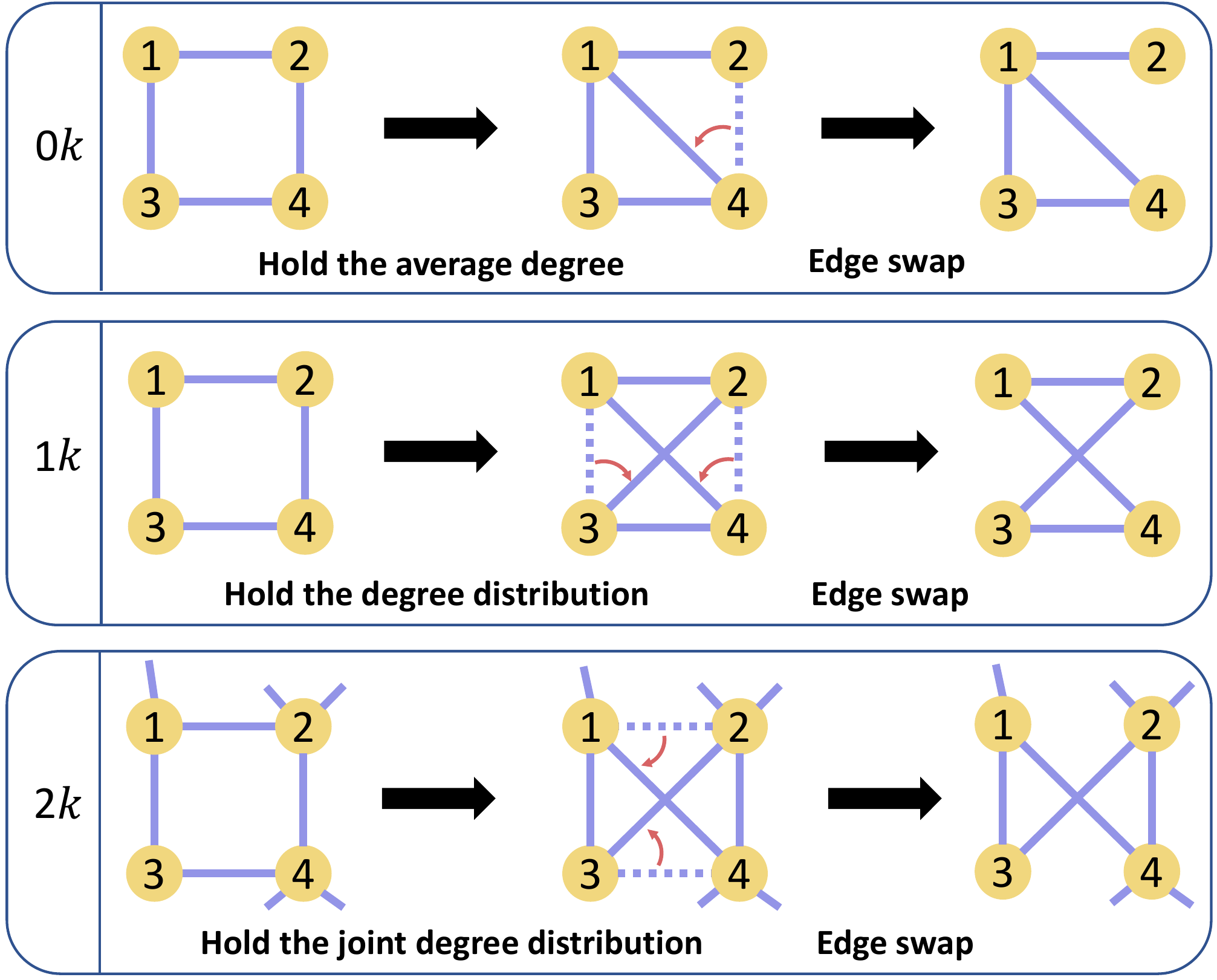}
  \caption{The augmentation methods of 0$k$, 1$k$, and 2$k$ null models. The legend that the red arrow points from the dotted line to the solid line represents a rewiring operation.}
  \label{fig:nullmodolaugmentation}
  \end{center}
\end{figure}

\subsection{Graph Data Augmentation}
Firstly, we use three classical null models to design the graph augmentation module, specifically the 0-order (0$k$), 1-order (1$k$), and 2-order (2$k$) null models. Then, we construct one standard null model-based heuristic graph augmentation strategy, Leaf Node Augmentation (LNA), and four approximate null model-based heuristic graph augmentation strategies, referred to as Approximate Data Augmentation (ADA), including Betweenness Centrality Augmentation (ADA-\textsc{Bc}), Clustering Coefficient Augmentation (ADA-\textsc{C}), Eigenvector Centrality Augmentation (ADA-\textsc{Ec}), and Closeness Centrality Augmentation (ADA-\textsc{Cc}). 

\begin{figure*}[!t]
  \begin{center}
  \includegraphics[width=\linewidth]{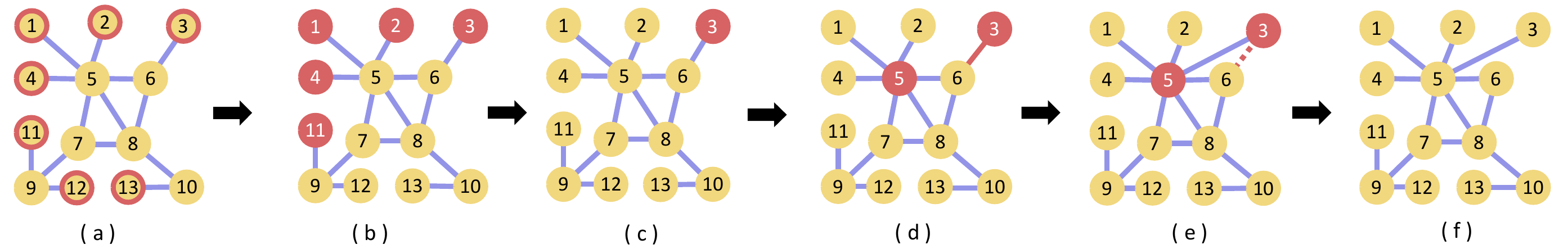}
  \caption{An example of LNA with augmentation cost coefficient $\alpha=0.2$. (a) Original graph; (b) The eligible leaf nodes marked by red in the graph; (c) The $\alpha*5$ augmented nodes ($v_3$ here) randomly selected from leaf nodes; (d) The nodes with the highest degree ($v_5$ here) among the neighbors of chosen leaf nodes; (e) The rewiring process, i.e., a new edge is established between $v_3$ and $v_5$, while that between $v_3$ and $v_6$ is disconnected; (f) The augmented graph $G_{aug}$.
}
  \label{fig:Leaf nodes augmentation of edge}
  \end{center}
\end{figure*}

\subsubsection{Standard Null Model-based Augmentation}
\textbf{Classical Null Model Augmentation.} Typically, there are two methods to generate null models. One is based on the configuration model\cite{ren2017time}, and the other is based on rewiring edges. In this paper, the method of rewiring edges is used to construct 0$k$, 1$k$, and 2$k$ null models. The null models with different orders hold different properties to remain consistent with the original graph. 
The corresponding augmentation strategies are briefly described in Fig. \ref{fig:nullmodolaugmentation}. 

The generation of 0$k$ null model is based on random rewiring, that is, randomly selecting an edge ($v_2$,$v_4$) to break and randomly selecting a pair of disconnected nodes ($v_1$,$v_4$) to connect. In general, the rewiring operation is performed for multiple times according to the experimental setting and the network scale in order to fully randomize the network. One can see that the 0$k$ augmentation model holds the same average degree as the original graph. 

As for the 1$k$ null model, its rewiring constraint is stricter than that of the 0$k$ null model. As shown by the 1$k$ null model example in Fig.~\ref{fig:nullmodolaugmentation}, one randomly break $(v_1,v_3)$, $(v_2,v_4)$ and then connect $(v_1,v_4)$, $(v_2,v_3)$ to keep the degree of each node unchanged before and after rewiring. That is, the 1$k$ null model will select two edges in each rewiring operation while maintaining the consistency of node degree distribution with the original graph on the basis of the 0$k$ null model. 

The augmentation strategy of higher-order null model is extended from that of the 1$k$ null model. Thus, the 2$k$ null model is obtained through adding a new restricted condition on the basis of the 1$k$ null model. As shown by the 2$k$ augmentation example in Fig. \ref{fig:nullmodolaugmentation}, rewiring is operated only when the nodes $v_2$ and $v_4$ (or $v_1$ and $v_3$) have the same degree, i.e., the degree values of the endpoints of the edges remain the same after rewiring. The augmentation distribution based on the 2$k$ null model is the joint degree distribution of the original graph. 


\textbf{Leaf Node Augmentation (LNA).} A leaf node has degree $1$, which is very common and significant in real-world networks, e.g., aldehyde, amino, methyl, and other functional groups on the benzene rings which can determine the chemical properties of the compounds~\cite{debnath1991structure}. Also, in the taxonomies of genes\cite{bioinformatics_btk048}, the leaf nodes have more important biological meanings than the internal nodes in some situation, for example filial samples are more important than parental samples in the study of genetic diseases with intergenerational genetic attributes.
The LNA is an augmentation strategy by fixing the proportion of leaf nodes. Given a graph $G = (V,E)$, denote the set of edges with leaf node by $E_{leaf} = \{(u_i,w_i)\in E|i=1,\cdots,p\}$, where $u_i$ is a leaf node in the graph $G$, $w_i$ is the neighbor of the leaf node $u_i$, and $p$ is the number of leaf nodes. LNA obtains augmentation graphs by rewiring the leaf nodes. The construction method is shown in Algorithm~\ref{alg:Leaf Nodes Algorithm}. To avoid generating more leaf nodes after rewiring, the edges in $E_{leaf}$ should be filtered by a filter $\mathcal{L}$ to get $E_{leaf}'=\{(u_i,w_i)|i=1,\cdots,q\}|(q<p)$, where the constraint on $w_i$ is that its degree remains more than $1$ after removing the leaf node $u_i$, with
\begin{equation}
  E_{leaf}'= \mathcal{L}(E_{leaf})
      =\bigcup^{p}_{i}{\{\mathbb{I}(u_i,w_i)\}},
  \label{eq:leaf edge filter}
\end{equation}
where the function $\mathbb{I}(u_i,w_i)$ is defined by,
\begin{equation}
\mathbb{I}(u_i,w_i)=
  \begin{cases}
(u_i,w_i), & d(w_i)>1 \\
\varnothing , & otherwise
  \end{cases}
  \label{eq:pi function}
\end{equation}
In order to ensure that no new leaf nodes are generated during the augmentation, each time an edge $(u_i,w_i)$ is chosen from $E_{leaf}'$, where the degree of $w_i$ must be subtracted by 1.
Then, randomly select $E_{del}\subset E_{leaf}'$ as the set of deleted edges, 
where $|E_{del}|=\alpha*|E_{leaf}'|$ and $\alpha$ is the cost coefficient of augmentation. When deleting the existing edges, the topology of the graph will be damaged to some extent. In order to make new leaf nodes carry as much information of their neighbors as possible, each $u_j$ will be reconnected to the node $\overline{w}_j$ with the highest degree among the neighbors of $w_j$.
Then, the set of adding edges is denoted as $E_{add}=\{(u_j,\overline{w}_j)|j=1,\cdots,\alpha*q\}$. Finally, based on the LNA, the original graph is modified to become a new graph $G_{aug}=(V,E')$, where 
\begin{equation}
  E' = E\cup E_{add}\verb|\|E_{del}.
  \label{eq:leaf nodes aug}
\end{equation}

Fig.~\ref{fig:Leaf nodes augmentation of edge} shows an example of LNA with augmentation cost coefficient $\alpha=0.2$. As is shown in Fig.~\ref{fig:Leaf nodes augmentation of edge} (a), there is a graph with seven leaf nodes $\{v_1, v_2, v_3, v_4, v_{11}, v_{12}, v_{13}\}$. Among them, leaf nodes $v_{11}, v_{12}$ have a common neighbor node $v_{9}$. If edges $(v_9, v_{11})$ and $(v_9, v_{12})$ are deleted at the same time, node $v_9$ will become a new leaf node, which will not meet Eq.~(\ref{eq:pi function}), so one of $v_{11}, v_{12}$ will be randomly selected to put into the eligible nodes set. Also, because $v_{10}$ will become a leaf node after deleting the edge $(v_{10}, v_{13})$, node $v_{13}$ does not meet the augmentation constraint. Thus, the five eligible leaf nodes $\{v_1, v_2, v_3, v_4, v_{11}\}$ are marked. Then, randomly select $\alpha*5$ eligible leaf nodes as augmented nodes, remove their original edges, and connect them to their 2-hop neighbor nodes with the highest degrees. Finally, the leaf node-based augmented graph is obtained. 

\begin{algorithm}[!t]
  \caption{Leaf Nodes Augmentation}
  \label{alg:Leaf Nodes Algorithm}
  \textbf{Input}: Original graph $G$ \\
  \textbf{Parameters}: Augmentation cost coefficient $\alpha$\\
  \textbf{Output}: Augmented graph $G_{aug}$
  \begin{algorithmic}[1]
  \STATE Get edges with leaf node $E_{leaf}$;
  \STATE Get $E'_{leaf}$ via Eq~(\ref{eq:leaf edge filter}) and Eq~(\ref{eq:pi function});
  \STATE $E_{del}\leftarrow$ {\rm RandomSample}($E'_{leaf},\alpha$);
  \FOR{$(u_j,w_j)\in E_{del}'$}
  \STATE $\overline{w}_j=\mathop{\arg\max}\limits_{v}{((G.{\rm neighbors}(w_j)).{\rm degree()})}$;
  \STATE $E_{add}.append((u_j,\overline{w}_j))$;
  \ENDFOR
  \STATE Get $G_{aug}$ via Eq.(\ref{eq:leaf nodes aug});
  \STATE Return $G_{aug}$;
  \end{algorithmic}
\end{algorithm}

\begin{figure}[!t]
  \begin{center}
  \includegraphics[width=\linewidth]{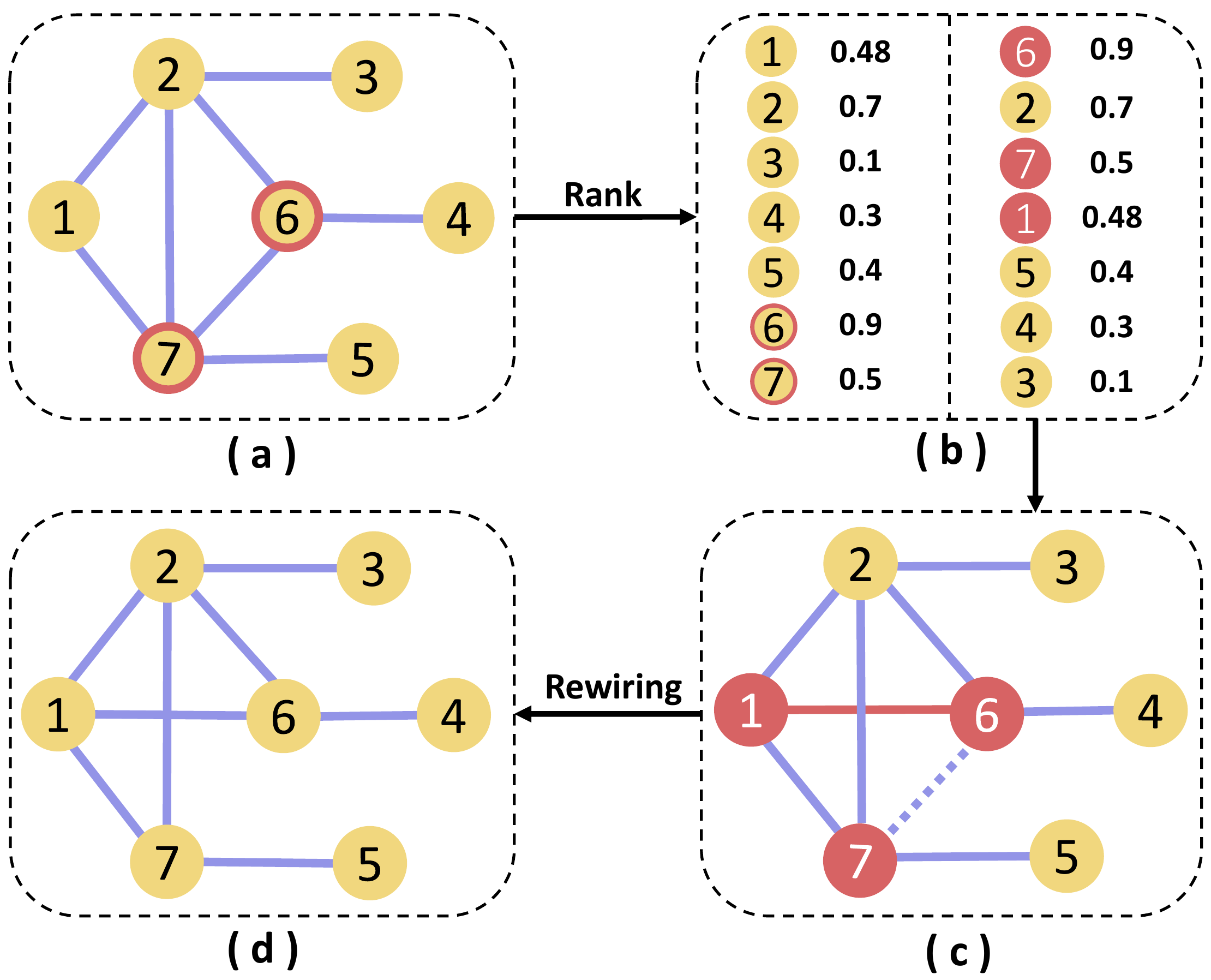}
  \caption{The process of approximate augmentation. (a) Original graph $G$, (b) features ranking and augmented node acquisition, (c) rewiring operation (the feature values are fictitious for illustration only), (d) augmented graph $G_{aug}$.}
  \label{fig:Approximate augmentation}
  \end{center}
\end{figure}

\begin{algorithm}[!t]
  \caption{Approxiamte Augmentation}
  \label{alg:Approximate augmentation Algorithm}
  \textbf{Input}: Original graph $G$ \\
  \textbf{Parameters}: Augmentation cost coefficient $\alpha$, Iterations $T$\\
  \textbf{Output}: Augmented graph $G_{aug}$
  \begin{algorithmic}[1]
  \STATE Initialize iteration = 0;
  \STATE Get the feature value of original graph $F$, the number of edges $m$, the number of nodes $n$;
  \FOR{iteration < $T$}
    \STATE Initialize swap = 0;
    \WHILE{swap < $\alpha*m$}
      \STATE Randomly sample an edge $e_1=(v_1,v_2)$ without leaf node;
      \STATE Get the set of feature values $\mathcal{S}$=$\{f_v|v=v_1,\cdots,v_n\}$;
      \STATE $u \leftarrow \mathop{\arg\max}\limits_{v}{(f_{v_1},f_{v_2})}$;
      \STATE $w \leftarrow \mathop{\arg\min}\limits_{v}{(f_{v_1},f_{v_2})}$;
      \STATE Get the nodes $\overline{w}$ that $f_{\overline{w}}$ is closest to $f_w$;
      \STATE $e_{del}\leftarrow e_1$;
      \STATE $e_{add}\leftarrow e_2={(u,\overline{w})}$;
      \STATE Rewiring to get the $G'$;
      \IF{$G'$ is connected}
        \STATE swap = swap + 1;
      \ELSE
        \STATE Cancel the rewiring;
      \ENDIF
    \ENDWHILE
  \ENDFOR
  \STATE Get $G_{aug}$ via Eq~(\ref{eq:approximate augmented2});
  \STATE Return $G_{aug}$;
  \end{algorithmic}
\end{algorithm}

\subsubsection{Approximate Null Model-based Augmentation}
\textbf{Approximate Data Augmentation (ADA).} Existing null models are based on simple features such as degree distribution, while centrality metrics based on network path structure, such as clustering coefficient, betweenness centrality, closeness centrality, or eigenvector centrality, not only consider the network topology but also summarize the participation or contribution of nodes to the network. It is not difficult to understand that, as the features become more complex, there will be more constraints in the generation of networks based on null models. Also, the more similar the generated network is to the original network, the more difficult the rewiring would become, i.e., fewer edges are appropriate to be selected to rewire under the stricter constraints. 
In order to address this, we propose a feature approximation method to reduce the impact of certain features in the augmentation process. Here, the concept of feature approximation is applied in the specific augmentation algorithms based on average clustering coefficient, average betweenness centrality, average closeness centrality, or average eigenvector centrality. In the following description, the graph feature values for a graph $G$ are denoted by the symbol $F$, and their feature values for a node $v$ are replaced with the symbol $f$.

As show in Fig.~\ref{fig:Approximate augmentation}, for a given graph $G=(V,E)$, we can first randomly select an edge without leaf node and set it as $e_1 = (v_6, v_7)$. Secondly, we get the list of feature values of each node in the graph $\mathcal{S}$=$\{f_v|v=v_1,\cdots,v_n\}$. Then, as shown in tables of Fig.~\ref{fig:Approximate augmentation} (b), we sort the list of feature values and get the node whose feature value is closest to that of $v_7$. Here, the node with the closest feature value is $v_1$. Meanwhile, we also do the operation of approximate augmentation by executing the equation in line 10 of Algorithm~\ref{alg:Approximate augmentation Algorithm} to get the node $v_1$ that has no edge with $v_6$ but with its feature value closest to the node $v_7$.
Next, we delete the edge ($v_6$, $v_7$) and connect node $v_6$ with $v_1$ to get the new edge $(v_6, v_1)$. After that, if the graph $G$ is connected, we perform rewiring as an effective augmentation operation; otherwise, cancel this rewiring operation and enter the next loop (line 14-19 in Algorithm~\ref{alg:Approximate augmentation Algorithm}). At this point, the augmentation operation for one edge has been completed, and it will happen $\alpha*m$ times in the process of graph augmentation to generate $G'$,
where $\alpha$ denotes the rewiring cost coefficient and $m$ denotes the number of edges in $G$. At the same time, we set the iteration parameter $T$ for this augmentation, and choose the best-augmented graph to return:
\begin{equation}
  G_{aug}= \mathop{\arg\min}\limits_{G'}{|F'-F|}.
  \label{eq:approximate augmented2}
\end{equation}
Thus, as long as $T$ is large enough, this model can generate an augmented graph $G_{aug}$ with high similarity. 

\section{Experiments}\label{sec:expset}
In this section, we conduct some experiments to evaluate the effectiveness of our graph data augmentation strategies for graph classification on a variety of real-world network datasets. We first introduce the datasets, then outline the graph classification methods adopted as baselines, and finally describle the experimental setup. After that, we show the experimental results with some discussion.

\subsection{Datasets} 
In order to access our augmentation methods, we adopt five commonly used benchmark datasets in experiments, which are BZR, COX2, MUTAG, OHSU, and ENZYMES. Among them, BZR, COX2, MUTAG, and ENZYMES are biological and chemical datasets, and OHSU is a brain dataset. The statistics of these datasets are summarized in Table~\ref{Tab:Datasets}.
\begin{itemize}
  \item \textbf{BZR}~\cite{Jeffrey2004Spline} is a dataset of 405 ligands for the benzodiazepine receptor, where the nodes and edges represent atoms and chemical bonds, respectively.
  They are classified to active and inactive compounds according to a predetermined threshold.
  \item \textbf{COX2}~\cite{Jeffrey2004Spline} is a dataset of 467 cyclooxygenase-2 with in vitro activity against human recombinant enzymes. Their graph topologies represent the position information of atoms and chemical bonds. They can also be classified into active and inactive compounds using a predetermined threshold.
  \item \textbf{MUTAG}~\cite{2012Subgraph} is a collection of nitroaromatic compounds, in which nodes and edges correspond to atoms and chemical bonds, respectively. Their classification is based on their mutagenicity to Salmonella typhimurium. 
  \item \textbf{OHSU}~\cite{2020Deep} composes the functional brain networks constructed from the whole brain fMRI map, where each node corresponds to a region of interest (ROI), and the edge represents the correlation between the two ROIs. The dataset can be divided to two classes based on the hyperactivity-impulsive attributes.
  \item \textbf{ENZYMES}~\cite{schomburg2004brenda} represents the protein macromolecules, where nodes indicate the secondary structure elements. If these nodes are neighbors along specific sequence, or they are neighbors in space within the protein structure, they will be connected. Each node is connected to its three nearest spatial neighbors. Further, according to different catalytic reactions, they can be divided into $6$ classes.
\end{itemize}


\begin{table}[!h]
  \centering
  \caption{Statistics of five datasets used in experiments. $N_G$, $N_C$, $\#Nodes$, and $\#Edges$ denote the number of graphs, the number of graph classes, the average number of nodes, and the average number of edges, respectively.}
  \begin{tabular}{l c c c c}
      \hline Datasets & $N_G$ & $N_C$ & $\#Nodes$ & $\#Edges$ \\
      \hline BZR & 405 & 2 & 43.75 & 38.44 \\
      COX2 & 467 & 2 & 41.22 & 43.45 \\
      MUTAG & 188 & 2 & 17.93 & 43.79 \\
      OHSU & 79 & 2 & 82.01 & 439.66 \\
      ENZYMES & 600 & 6 & 32.63 & 62.14 \\
      \hline
  \end{tabular}
  \label{Tab:Datasets}
\end{table}

\subsection{Graph Classification Methods}
Here, we utilize some graph classification methods to learn the representations of the original and augmented graphs and then predict the class of the given graph. Under the present framework, we adopt five different methods namely SF, NetLSD, gl2vec, Graph2vec, and Diffpool, where SF and Graph2vec are graph embedding methods, NetLSD and gl2vec are graph kernel models, and Diffpool is an end-to-end graph neural network method.
\begin{itemize}
  \item \textbf{SF}~\cite{de2018simple} is an embedding method, which relies on spectral features of the graph. It performs graph classification based on the spectral decomposition of the graph Laplacian.
  \item \textbf{Graph2vec}~\cite{narayanan2017graph2vec} is the first unsupervised embedding approach for an entire network, which can learn data-driven distributed representations of arbitrary sized graphs.
  \item \textbf{NetLSD}~\cite{tsitsulin2018netlsd} is a graph kernel model, which compares graphs and achieves graph classification by extracting a compact signature that inherits the formal properties of the Laplacian spectrum.
  \item \textbf{gl2vec}~\cite{tu2019gl2vec} generates feature representation by static or temporal network graphlet distribution and a null model to compare with random graphs. 
  \item \textbf{Diffpool}~\cite{2018Hierarchical} introduces a way to aggregate nodes to learn a graph representation that contains hierarchical information. It can be combined with GNN architectures in an end-to-end fashion.
\end{itemize}

\begin{table*}[!t]
  \centering
  \caption{Graph classification results of original and standard null model-based augmentation models. The best results are marked in bold.}
  \resizebox{2\columnwidth}{!}{
    \begin{tabular}{c|c|cccc|cccc|cccc|cccc|c|c}
    \toprule
    \multirow{2}[4]{*}{Datasets} & \multirow{2}[4]{*}{Aug Model} & \multicolumn{4}{c|}{SF}       & \multicolumn{4}{c|}{Graph2vec} & \multicolumn{4}{c|}{NetLSD}   & \multicolumn{4}{c|}{gl2vec}   & \multirow{2}[4]{*}{Diffpool} & \multirow{2}[4]{*}{$R_{Gain}$} \\
\cmidrule{3-18}          &       & SVM & Logistic & KNN   & RF    & SVM & Logistic & KNN   & RF    & SVM & Logistic & KNN   & RF    & SVM & Logistic & KNN   & RF    &       &  \\
    \midrule
    \multirow{5}[2]{*}{BZR} & orignal & 0.796 & 0.734 & 0.805 & 0.838 & 0.799 & 0.799 & 0.837 & 0.839 & 0.807 & 0.757 & 0.803 & 0.819 & 0.799 & 0.807 & 0.843 & 0.834 & 0.827 & / \\
          & 0$k$    & 0.799 & \textbf{0.811} & 0.809 & 0.841 & 0.804 & \textbf{0.837} & 0.852 & 0.846 & 0.813 & 0.81  & 0.81  & 0.833 & 0.801 & 0.821 & 0.851 & \textbf{0.839} & 0.846 & 2.11\% \\
          & 1$k$    & 0.806 & 0.802 & 0.811 & 0.841 & 0.802 & 0.833 & 0.856 & 0.839 & 0.816 & 0.81  & 0.815 & 0.833 & 0.804 & \textbf{0.842} & \textbf{0.857} & 0.834 & \textbf{0.853} & 2.33\% \\
          & 2$k$    & \textbf{0.807} & 0.809 & 0.807 & \textbf{0.843} & \textbf{0.804} & 0.836 & \textbf{0.859} & \textbf{0.847} & \textbf{0.821} & 0.809 & \textbf{0.815} & \textbf{0.834} & \textbf{0.804} & 0.841 & 0.853 & 0.835 & 0.851 & \textbf{2.49\%} \\
          & LNA   & 0.806 & 0.808 & \textbf{0.816} & 0.842 & 0.801 & 0.833 & 0.849 & 0.847 & 0.815 & \textbf{0.812} & 0.814 & 0.832 & 0.803 & 0.831 & 0.852 & 0.838 & 0.849 & 2.29\% \\
    \midrule
    \multirow{5}[2]{*}{COX2} & orignal & 0.777 & 0.735 & 0.779 & 0.77  & 0.778 & 0.746 & 0.786 & 0.783 & 0.77  & 0.666 & 0.774 & 0.752 & 0.777 & 0.728 & 0.782 & 0.788 & 0.804 & / \\
          & 0$k$    & 0.78  & 0.782 & 0.783 & 0.783 & 0.779 & 0.789 & 0.788 & 0.791 & 0.776 & 0.785 & 0.78  & 0.768 & 0.778 & 0.789 & 0.787 & 0.79  & 0.823 & 2.91\% \\
          & 1$k$    & \textbf{0.782} & \textbf{0.786} & \textbf{0.796} & \textbf{0.79} & \textbf{0.783} & \textbf{0.809} & 0.791 & \textbf{0.803} & 0.776 & \textbf{0.786} & 0.793 & 0.773 & 0.779 & \textbf{0.799} & 0.797 & 0.794 & \textbf{0.83} & \textbf{3.80\%} \\
          & 2$k$    & 0.781 & 0.781 & 0.794 & 0.786 & 0.781 & 0.799 & \textbf{0.8} & 0.8   & \textbf{0.784} & 0.786 & \textbf{0.795} & \textbf{0.778} & \textbf{0.782} & 0.797 & \textbf{0.799} & \textbf{0.796} & 0.822 & 3.75\% \\
          & LNA   & 0.778 & 0.785 & 0.785 & 0.782 & 0.78  & 0.795 & 0.787 & 0.787 & 0.782 & 0.784 & 0.785 & 0.777 & 0.779 & 0.794 & 0.784 & 0.791 & 0.828 & 3.16\% \\
    \midrule
    \multirow{5}[2]{*}{MUTAG} & orignal & 0.822 & 0.824 & 0.829 & 0.854 & 0.744 & 0.777 & 0.772 & 0.818 & 0.823 & 0.786 & 0.827 & 0.837 & 0.741 & 0.795 & 0.781 & 0.797 & 0.759 & / \\
          & 0$k$    & 0.835 & \textbf{0.864} & 0.837 & \textbf{0.871} & 0.746 & 0.831 & 0.781 & 0.848 & 0.843 & 0.857 & 0.855 & 0.864 & 0.747 & \textbf{0.843} & 0.809 & 0.836 & 0.835 & 3.82\% \\
          & 1$k$    & 0.834 & 0.853 & 0.845 & 0.862 & 0.742 & \textbf{0.837} & \textbf{0.817} & 0.849 & \textbf{0.846} & \textbf{0.864} & \textbf{0.859} & \textbf{0.871} & \textbf{0.749} & 0.839 & \textbf{0.84} & 0.817 & \textbf{0.853} & \textbf{4.39\%} \\
          & 2$k$    & 0.836 & 0.858 & 0.844 & 0.863 & 0.746 & 0.824 & 0.816 & 0.835 & 0.845 & 0.863 & 0.855 & 0.865 & 0.743 & 0.82  & 0.828 & 0.828 & 0.853 & 3.98\% \\
          & LNA   & \textbf{0.839} & 0.858 & \textbf{0.856} & 0.865 & \textbf{0.747} & 0.825 & 0.805 & \textbf{0.863} & 0.836 & 0.849 & 0.854 & 0.86  & 0.747 & 0.84  & 0.833 & \textbf{0.838} & 0.851 & 4.30\% \\
    \midrule
    \multirow{5}[2]{*}{OHSU} & orignal & 0.61  & 0.565 & 0.61  & 0.639 & 0.557 & 0.58  & 0.577 & 0.582 & 0.547 & 0.504 & 0.55  & 0.558 & 0.557 & 0.535 & 0.542 & 0.516 & 0.476 & / \\
          & 0$k$    & 0.678 & 0.643 & \textbf{0.646} & \textbf{0.735} & 0.557 & 0.635 & 0.603 & 0.628 & 0.603 & 0.552 & 0.615 & 0.654 & 0.557 & 0.587 & 0.577 & 0.62  & 0.585 & 10.34\% \\
          & 1$k$    & 0.683 & 0.645 & 0.638 & 0.714 & 0.559 & \textbf{0.686} & 0.627 & 0.678 & 0.661 & 0.569 & 0.604 & \textbf{0.688} & 0.614 & 0.608 & \textbf{0.668} & 0.63  & 0.588 & 14.46\% \\
          & 2$k$    & \textbf{0.686} & 0.628 & 0.64  & 0.732 & \textbf{0.562} & 0.635 & \textbf{0.662} & 0.676 & \textbf{0.663} & 0.566 & \textbf{0.636} & 0.66  & \textbf{0.614} & 0.587 & 0.639 & \textbf{0.68} & 0.606 & \textbf{14.63\%} \\
          & LNA   & 0.678 & \textbf{0.658} & 0.64  & 0.714 & 0.557 & 0.648 & 0.625 & \textbf{0.683} & 0.656 & \textbf{0.584} & 0.622 & 0.68  & 0.564 & \textbf{0.63} & 0.614 & 0.655 & \textbf{0.626} & 14.30\% \\
    \midrule
    \multirow{5}[2]{*}{ENZYMES} & orignal & 0.312 & 0.241 & 0.274 & 0.385 & 0.365 & 0.244 & 0.28  & 0.322 & 0.34  & 0.202 & 0.311 & 0.343 & 0.356 & 0.245 & 0.273 & 0.304 & 0.353 & / \\
          & 0$k$    & 0.324 & 0.266 & 0.292 & 0.407 & 0.309 & 0.275 & 0.302 & 0.353 & 0.357 & 0.254 & 0.337 & 0.371 & 0.291 & 0.284 & 0.278 & 0.348 & 0.357 & 6.10\% \\
          & 1$k$    & \textbf{0.343} & \textbf{0.273} & \textbf{0.335} & 0.429 & \textbf{0.399} & 0.28  & 0.334 & 0.369 & 0.363 & 0.26  & 0.355 & 0.377 & 0.374 & 0.285 & 0.294 & 0.361 & 0.427 & 14.30\% \\
          & 2$k$    & 0.336 & 0.269 & 0.319 & 0.425 & 0.398 & \textbf{0.288} & \textbf{0.346} & \textbf{0.38} & 0.362 & \textbf{0.267} & 0.348 & 0.381 & \textbf{0.383} & \textbf{0.29} & 0.295 & \textbf{0.368} & \textbf{0.437} & \textbf{15.02\%} \\
          & LNA   & 0.336 & 0.263 & 0.333 & \textbf{0.439} & 0.377 & 0.261 & 0.321 & 0.359 & \textbf{0.39} & 0.262 & \textbf{0.369} & \textbf{0.382} & 0.358 & 0.27  & \textbf{0.298} & 0.347 & 0.422 & 12.75\% \\
    \bottomrule
    \end{tabular}%
    
  }
  \label{Tab:NullModel_compare}%
\end{table*}%

\begin{table*}[!t]
  \centering
  \caption{Graph classification results of original and approximate augmentation models. The best results are marked in bold.}
  \resizebox{2\columnwidth}{!}{
    \begin{tabular}{c|c|cccc|cccc|cccc|cccc|c|c}
    \toprule
    \multirow{2}[4]{*}{Datasets} & \multirow{2}[4]{*}{Aug Model} & \multicolumn{4}{c|}{SF}       & \multicolumn{4}{c|}{Graph2vec} & \multicolumn{4}{c|}{NetLSD}   & \multicolumn{4}{c|}{gl2vec}   & \multirow{2}[4]{*}{Diffpool} & \multirow{2}[4]{*}{$R_{Gain}$} \\
\cmidrule{3-18}          &       & SVM & Logistic & KNN   & RF    & SVM & Logistic & KNN   & RF    & SVM & Logistic & KNN   & RF    & SVM & Logistic & KNN   & RF    &       &  \\
    \midrule
    \multirow{5}[2]{*}{BZR} & orignal & 0.796 & 0.734 & 0.805 & 0.838 & 0.799 & 0.799 & 0.837 & 0.839 & 0.807 & 0.757 & 0.803 & 0.819 & 0.799 & 0.807 & 0.843 & 0.834 & 0.827 & / \\
          & ADA-C & 0.802 & 0.808 & 0.808 & 0.84  & 0.801 & 0.826 & 0.846 & 0.842 & 0.81  & 0.808 & 0.809 & 0.825 & 0.799 & 0.827 & 0.846 & \textbf{0.844} & 0.846 & 1.85\% \\
          & ADA-\textsc{Bc} & 0.803 & 0.81  & \textbf{0.812} & 0.843 & 0.8   & 0.832 & 0.847 & 0.841 & 0.809 & 0.804 & 0.81  & 0.83  & \textbf{0.803} & 0.839 & 0.848 & 0.837 & 0.848 & 2.06\% \\
          & ADA-\textsc{Cc} & \textbf{0.804} & 0.809 & 0.811 & \textbf{0.844} & 0.803 & 0.825 & 0.847 & \textbf{0.844} & \textbf{0.812} & 0.806 & \textbf{0.814} & \textbf{0.833} & 0.801 & 0.831 & \textbf{0.854} & 0.84  & 0.848 & 2.13\% \\
          & ADA-\textsc{Ec} & 0.804 & \textbf{0.81} & 0.802 & 0.843 & \textbf{0.804} & \textbf{0.832} & \textbf{0.854} & 0.839 & 0.812 & \textbf{0.816} & 0.809 & 0.829 & 0.798 & \textbf{0.841} & 0.849 & 0.841 & \textbf{0.852} & \textbf{2.20\%} \\
    \midrule
    \multirow{5}[2]{*}{COX2} & orignal & 0.777 & 0.735 & 0.779 & 0.77  & 0.778 & 0.746 & 0.786 & 0.783 & 0.77  & 0.666 & 0.774 & 0.752 & 0.777 & 0.728 & 0.782 & 0.788 & 0.804 & / \\
          & ADA-C & 0.78  & 0.785 & 0.782 & 0.78  & 0.778 & 0.78  & 0.783 & 0.792 & 0.773 & 0.779 & 0.777 & 0.764 & 0.778 & 0.784 & 0.79  & 0.793 & 0.822 & 2.66\% \\
          & ADA-\textsc{Bc} & \textbf{0.783} & 0.783 & \textbf{0.788} & \textbf{0.783} & 0.779 & \textbf{0.797} & 0.788 & \textbf{0.797} & \textbf{0.78} & \textbf{0.785} & \textbf{0.795} & \textbf{0.776} & \textbf{0.779} & 0.787 & 0.786 & \textbf{0.793} & \textbf{0.827} & \textbf{3.33\%} \\
          & ADA-\textsc{Cc} & 0.78  & 0.779 & 0.782 & 0.779 & \textbf{0.78} & 0.797 & 0.787 & 0.792 & 0.778 & 0.784 & 0.778 & 0.77  & 0.779 & \textbf{0.795} & 0.786 & 0.792 & 0.824 & 3.00\% \\
          & ADA-\textsc{Ec} & 0.78  & \textbf{0.789} & 0.781 & 0.78  & 0.778 & 0.785 & \textbf{0.792} & 0.797 & 0.774 & 0.779 & 0.776 & 0.771 & 0.778 & 0.782 & \textbf{0.791} & 0.79  & 0.821 & 2.85\% \\
    \midrule
    \multirow{5}[2]{*}{MUTAG} & orignal & 0.822 & 0.824 & 0.829 & 0.854 & 0.744 & 0.777 & 0.772 & 0.818 & 0.823 & 0.786 & 0.827 & 0.837 & 0.741 & 0.795 & 0.781 & 0.797 & 0.759 & / \\
          & ADA-C & 0.837 & 0.845 & 0.847 & 0.874 & 0.745 & 0.807 & 0.792 & 0.823 & 0.845 & 0.856 & 0.856 & 0.861 & \textbf{0.751} & 0.815 & 0.784 & 0.801 & 0.837 & 2.89\% \\
          & ADA-\textsc{Bc} & \textbf{0.841} & 0.863 & 0.844 & \textbf{0.876} & \textbf{0.746} & 0.807 & 0.78  & 0.835 & \textbf{0.854} & \textbf{0.868} & 0.853 & 0.866 & 0.748 & \textbf{0.843} & 0.809 & \textbf{0.825} & 0.84  & 3.78\% \\
          & ADA-\textsc{Cc} & 0.837 & \textbf{0.866} & \textbf{0.851} & 0.87  & 0.745 & 0.818 & \textbf{0.803} & \textbf{0.841} & 0.842 & 0.867 & 0.854 & 0.869 & 0.748 & 0.839 & \textbf{0.826} & 0.819 & \textbf{0.844} & \textbf{4.09\%} \\
          & ADA-\textsc{Ec} & 0.833 & 0.839 & 0.839 & 0.868 & 0.746 & \textbf{0.823} & 0.8   & 0.835 & 0.849 & 0.859 & \textbf{0.857} & \textbf{0.873} & 0.749 & 0.825 & 0.804 & 0.809 & 0.836 & 3.40\% \\
    \midrule
    \multirow{5}[2]{*}{OHSU} & orignal & 0.61  & 0.565 & 0.61  & 0.639 & 0.557 & 0.58  & 0.577 & 0.582 & 0.547 & 0.504 & 0.55  & 0.558 & 0.557 & 0.535 & 0.542 & 0.516 & 0.476 & / \\
          & ADA-C & 0.671 & 0.628 & \textbf{0.646} & 0.696 & 0.557 & \textbf{0.651} & 0.604 & 0.62  & 0.591 & 0.562 & 0.583 & 0.639 & 0.557 & \textbf{0.641} & 0.626 & 0.62  & 0.584 & 10.45\% \\
          & ADA-\textsc{Bc} & 0.684 & 0.627 & 0.638 & 0.699 & 0.557 & 0.643 & 0.635 & \textbf{0.646} & \textbf{0.623} & 0.567 & 0.605 & 0.628 & 0.557 & 0.618 & 0.588 & 0.633 & 0.573 & 10.88\% \\
          & ADA-\textsc{Cc} & \textbf{0.686} & 0.617 & 0.646 & \textbf{0.709} & 0.557 & 0.629 & \textbf{0.647} & 0.617 & 0.612 & 0.562 & 0.617 & 0.635 & 0.557 & 0.584 & \textbf{0.627} & 0.622 & 0.588 & 10.79\% \\
          & ADA-\textsc{Ec} & 0.656 & \textbf{0.64} & 0.643 & 0.702 & \textbf{0.557} & 0.613 & 0.6   & 0.628 & 0.582 & \textbf{0.57} & \textbf{0.637} & \textbf{0.653} & \textbf{0.557} & 0.623 & 0.605 & \textbf{0.661} & \textbf{0.589} & \textbf{10.94\%} \\
    \midrule
    \multirow{5}[2]{*}{ENZYMES} & orignal & 0.312 & 0.241 & 0.274 & 0.385 & \textbf{0.365} & 0.244 & 0.28  & 0.322 & 0.34  & 0.202 & 0.311 & 0.343 & \textbf{0.356} & 0.245 & 0.273 & 0.304 & 0.353 & / \\
          & ADA-C & \textbf{0.329} & 0.255 & 0.297 & 0.41  & 0.319 & 0.28  & 0.301 & 0.344 & 0.357 & \textbf{0.252} & \textbf{0.344} & 0.366 & 0.299 & 0.283 & 0.277 & 0.343 & \textbf{0.419} & \textbf{7.19\%} \\
          & ADA-\textsc{Bc} & 0.328 & 0.262 & 0.297 & 0.407 & 0.316 & 0.278 & 0.301 & 0.339 & \textbf{0.36} & 0.243 & 0.341 & 0.371 & 0.291 & 0.283 & \textbf{0.281} & 0.33  & 0.405 & 6.39\% \\
          & ADA-\textsc{Cc} & 0.329 & \textbf{0.27} & 0.301 & \textbf{0.412} & 0.323 & \textbf{0.282} & 0.298 & 0.343 & 0.355 & 0.238 & 0.342 & 0.371 & 0.294 & \textbf{0.293} & 0.27  & 0.339 & 0.416 & 7.19\% \\
          & ADA-\textsc{Ec} & 0.326 & 0.264 & \textbf{0.301} & 0.405 & 0.314 & 0.269 & \textbf{0.31} & \textbf{0.345} & 0.358 & 0.238 & 0.343 & \textbf{0.373} & 0.295 & 0.284 & 0.274 & \textbf{0.349} & 0.417 & 6.91\% \\
    \bottomrule
    \end{tabular}%
  }
  \label{Tab:similar_compare}%
\end{table*}%

\subsection{Experimental Setup\label{Experimental Framework}}
In this study, the embedding dimension of all graph kernels and embedding are set as $128$. For \emph{SF}, the random seed value is set to 42. For \emph{Graph2vec} and \emph{gl2vec}, the number of cores is set to 4. Given that these methods are based on the rooted subgraphs, some parameters are related to the setting of the WL kernel, where the number of Weisfeiler-Lehman iterations is 2. Also, the parameters are set to commmonly-used values: the learning rate is set to $0.025$, the epochs is set to $500$. For \emph{NetLSD}, the scheme calculates the heat kernel trace of the normalized Laplacian matrix over a vector of time scales. If the matrix is large, it switches to an approximation of the eigenvalues. Specifically, the number of eigenvalue approximations is set to $200$, whereas the minimum and maximum time scale interval are set to $-2.0$ and $2.0$, respectively. For \emph{Diffpool}, the parameter epochs are set to $3000$ and the other parameters are set to default values~\cite{2018Hierarchical}. 
In addition, the first four unsupervised representation methods, SF, Graph2vec, NetLSD, and gl2vec, are paired with four machine learning algorithms to implement the graph classification task, where the four machine learning algorithms are Support Vector Machine classifier based on radial basis kernel (\emph{SVM}), Logistic regression classifier (\emph{Logistic}), K-Nearest Neighbors classifier (\emph{KNN}) and Random Forest classifier (\emph{RF}). Therefore, there are totally $4\times{4}+1=17$ kinds of graph classification schemes in validation experiments.

Considering that lower augmentation cost coefficient will make the features value closer to the original value, we set the modified edge connection ratio (augmentation cost coefficient) of each augmentation model to $\alpha$=$0.2$ and set the number of the iteration of approximate augmentation as $T$=$5$. In our experiments, each dataset are divided into the training set, validation set, and test set with the ratio of 7:1:2, where the training set is augmented by the augmentation strategies developed in this paper. Then, we use the data filtering method in~\cite{zhou2020m} to filter the augmented graph set. Finally, we feed the augmented training set into the different graph classification classifiers for training. 

In the experiments, the following metrics are adopted to evaluate the graph classification performance of different augmentation strategies:
\begin{itemize}
  \item \textbf{Accuracy}. Accuracy measures the classification performance with the proportion of correctly classified graphs over all graphs in the dataset.

  \item \textbf{Success Rate}. The augmentation success rate refers to the ratio of the cases in which the augmented classification accuracy is higher than the original classification accuracy to the total.

  \item \textbf{Relative Gain Ratio}. The relative gain Ratio is defined as:
  \begin{equation}
    R_{gain}=\frac{Acc_{aug}-Acc_{ori}}{Acc_{ori}}\times 100\% ,
    \label{Gain Coefficient}
  \end{equation}
  where $Acc_{aug}$ and $Acc_{ori}$ denote the augmented and original classification accuracy, respectively. 
\end{itemize}

\begin{figure}[!t]
  \begin{center}
  \includegraphics[width=1\linewidth]{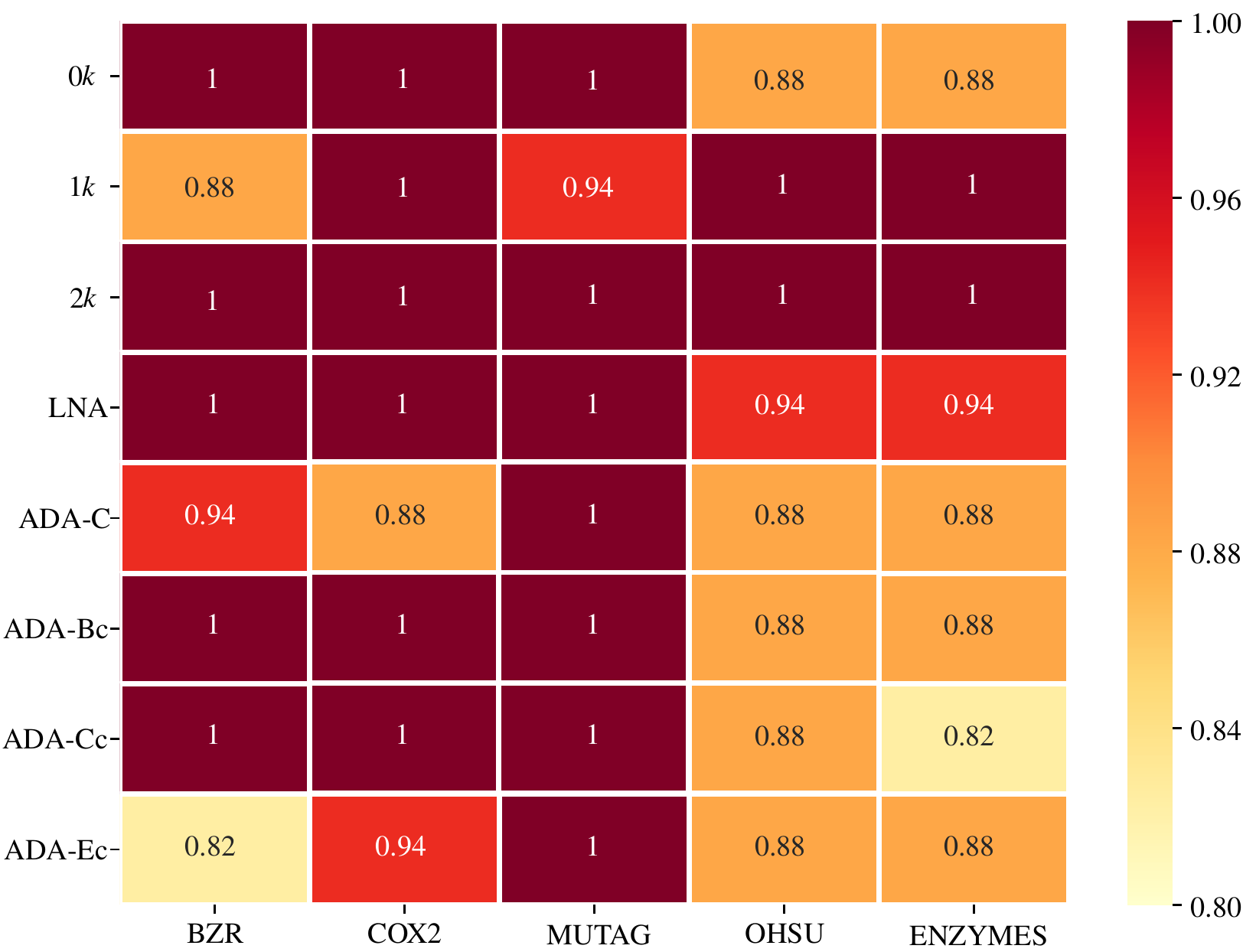}
  \caption{Statistics of augmentation success rate of various datasets.}
  \label{fig:Success Rate}
  \end{center}
\end{figure}


\subsection{Results and Discussion}\label{sec:result}
The experimental results conducted on the five datasets with the configuration in Section~\ref{sec:expset} are summarized in Table~\ref{Tab:NullModel_compare} and Table~\ref{Tab:similar_compare}. The data in the two tables consist of the classification accuracy of the original model and the augmentation model, and the average of the relative gain ratio of each augmentation method in different classification mechanisms.
The data are analyzed from different perspectives such as the performance of augmentation models, the augmentation success rate, and the analysis of time complexity.

\subsubsection{Performance of Augmentation Models} 
Table~\ref{Tab:NullModel_compare} presents the graph classification results of the original and standard null model-based augmentation models. It is easy to see that, 0$k$, 1$k$, 2$k$, and LNA augmentation models significantly improve the performance of graph classification compared with the original models based on the five graph representation methods, where the 2$k$ augmentation model based on \emph{gl2vec-RF} for the OHSU dataset even leads to an improvement of 16.4\%. It is worth noting that both 1$k$ and 2$k$ null model-based augmentation strategies generally have the best augmentation effectiveness on all datasets, which achieve the average gain ratio of 14.46\% and 14.63\% respectively for OHSU, and 14.30\% and 15.02\% respectively for ENZYMES. This is reasonable because these two augmentation models maintain the basic features such as (joint) degree distribution, which can better describe the network correlation.

Also, due to the approximate nature of the ADA methods, extra bias could be introduced, and thus these methods are typically somewhat weaker in enhancing graph classification models, as shown in Table~\ref{Tab:similar_compare}. As can be seen, the best performance is achieved with average gain ratio of 10.94\% (obtained by ADA-\textsc{Ec} for OHSU), which is about 4\% less than the 2$k$ null model-based augmentation model. For the multi-class dataset ENZYMES, the classification results obtained by the standard null model-based augmentation models are also significantly better compared with ADA models.
Besides, it is found that better classification enhancement performance generally occurs with the \emph{Diffpool} method or unsupervised representation methods with \emph{RF} classifier, which indicates that the effectiveness of these data augmentation methods could be further improved by designing appropriate graph representation and classification methods.

Moreover, we also investigate the augmentation success rate based on different augmentation models 
using the five datasets. According to Table~\ref{Tab:NullModel_compare} and Table~\ref{Tab:similar_compare}, one can see that, the augmentation models outperform the original model in 168 out of 170 cases. Also, the augmentation success rate of the 2$k$ is higher than other augmentation models, even reaching 100\% on the five datasets, as shown in Fig.~\ref{fig:Success Rate}. The minimum augmentation success rates of ADA-\textsc{Cc} and ADA-\textsc{Ec} also reach 82\%. Overall, the average augmented success rate of all the augmentation models is higher than 91\% on these datasets. This phenomenon supports the conclusion that our augmentation methods can indeed improve the performance of graph classification.

\begin{figure*}[!t]
  \begin{center}
  \includegraphics[width=1.025\linewidth]{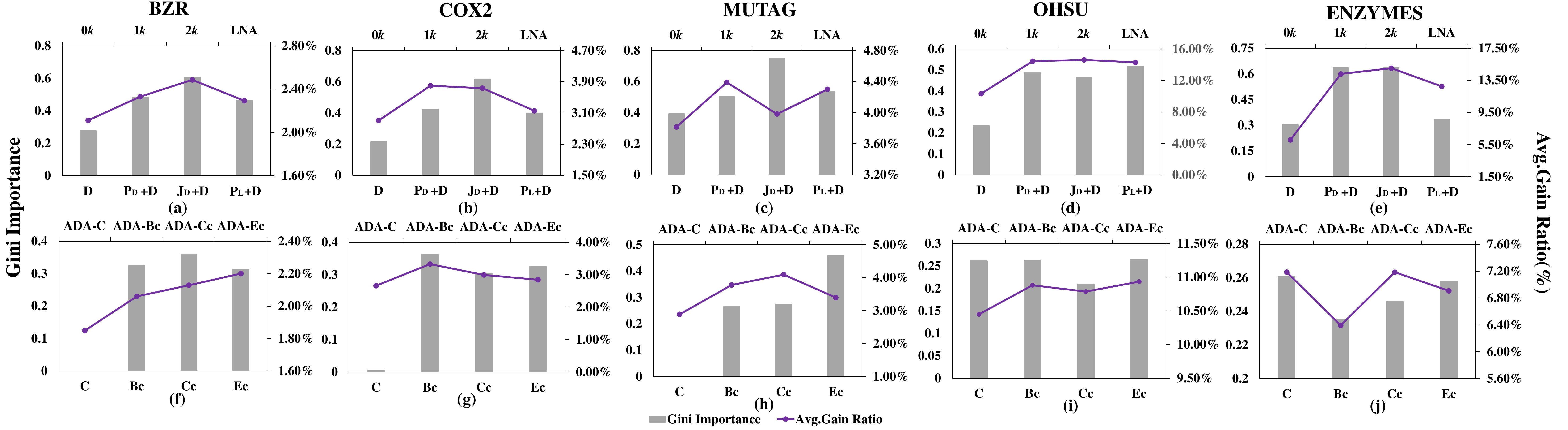}
  \caption{The compound chart of the \emph{Gini Importance} of features and $Avg.Gain\ Ratio$, where the features are the average degree of graph ($D$), the degree distribution of graph ($P_D$), the joint degree distribution of graph ($J_D$), the proportion of leaf nodes of graph ($P_L$), the clustering coefficient of graph($C$), the betweenness centrality of graph ($B_C$), the closeness centrality of graph ($C_C$), and the eigenvector centrality of graph ($E_C$).}
  \label{fig:features}
  \end{center}
\end{figure*}

\begin{figure*}[!t]
  \begin{center}
  \includegraphics[width=1\linewidth]{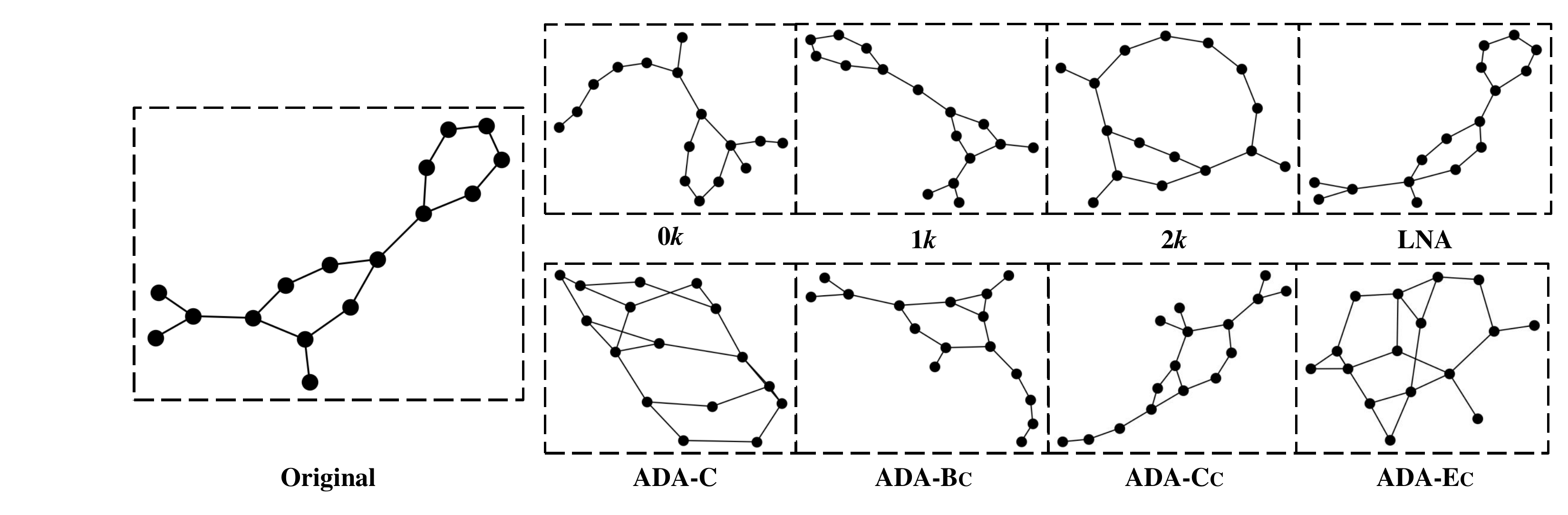}
  \caption{The seventh graph in MUTAG and its augmentation graphs. Obviously, compared with other augmentation graphs, the 2$k$, ADA-C, ADA-\textsc{Bc}, ADA-\textsc{Cc}, and ADA-\textsc{Ec} augmented graphs severely damaged the benzene ring of the original graph.}
  \label{fig:graph_compare}
  \end{center}
\end{figure*}

\begin{table}[!t]
  \centering
  \caption{Average augmentation time for the graph on five datasets (unit: second).}
  \begin{tabular}{cccccc}
    \toprule
    Dataset & BZR   & COX2  & MUTAG & OHSU  & ENZYMES \\
    \midrule
    0$k$    & 1.087  & 0.674  & 0.012  & 0.777  & 0.240  \\
    1$k$    & 0.006  & 0.004  & 0.003  & 0.066  & 0.002  \\
    2$k$    & 0.006  & 0.008  & 0.002  & 0.066  & 0.002  \\
    LNA   & 0.009  & 0.011  & 0.004  & 0.088  & 0.003  \\
    ADA-C & 0.072  & 0.113  & 0.036  & 3.673  & 0.138  \\
    ADA-BC & 0.969  & 3.388  & 0.216  & 46.831  & 1.478  \\
    ADA-CC & 0.457  & 1.771  & 0.143  & 18.363  & 0.751  \\
    ADA-EC & 0.763  & 3.158  & 0.392  & 3.801  & 1.102  \\
    \bottomrule
    \end{tabular}%
  \label{tab:runtime}%
\end{table}%

\subsubsection{Design of Null Model-Based Augmentation Models} 
Since our augmentation models show general improvement on the five datasets, we further dissect and illustrate their different performances on some particular graphs. We extract graph features manually and utilize \emph{Gini importance} computed by the \emph{Random Forest} \cite{1522531,breiman2001random,menze2009comparison} classifier to evaluate the feature importance.
The feature with the high \emph{Gini importance} plays a more significant role than others in graph classification. Fig.~\ref{fig:features} shows a compound chart consisting of the \emph{Gini importance} (bar) of features and the average gain ratio (line) of augmentation models,
where (a)-(e) are based on the null model-based augmentation, and (f)-(j) are based on the approximate null model-based augmentation. Interestingly, we find the consistency between the augmentation effects of the augmentation models and their corresponding features. 
This phenomenon is more prominent in the standard null model-based augmentation. 
Also, we find that the 2$k$ augmentation model on MUTAG dataset has high \emph{Gini importance} but relatively low average gain ratio. A possible reason is that the augmentation based on 2$k$ null model will break the structure of the benzene ring, which however is important for the classification of MUTAG. The non-positive trends in Fig~\ref{fig:features} (f)-(j) are also expected. Although the approximate augmentation methods are designed based on the principle of ensuring the consistency of the features as much as possible, its randomness will inevitably bring bias. Therefore, it is worth exploring the possibility to design particular augmentation strategies for different kinds of network structures. 

As an example, we visualize the different structures generated by the eight null model augmentation models on the seventh graph from the MUTAG dataset, as shown in Fig~\ref{fig:graph_compare}. MUTAG is a dataset of nitroaromatic compounds. After augmentation, one can see that the augmented graphs generated by different models have quite different structures. Compared with null-model augmented graphs, the approximate augmented graphs have less structural similarity to the original network. In particular, by adopting 2$k$, ADA-C, ADA-\textsc{Ec}, we find that not only the nitroaromatic structure is destroyed, but also the reconstructed six-membered ring may not be benzene rings composed of carbon. This also partly explains the phenomenon that in many cases the approximate augmentation methods are less effective than the standard null model-based augmentation methods.

Indeed, the main purpose of using the null model is to maintain the non-trivial features of a graph and gradually approximate the original graph. The results in Table \ref{Tab:NullModel_compare} and Table \ref{Tab:similar_compare} suggest that the augmentation methods are effective on both two-class and multi-class datasets, the reason being that the key feature of two-class datasets may be single, making two-class datasets easy to classify. However, the classification standard for the multi-class datasets could be different intervals of a key feature. This also gives us some inspiration to build diverse null models for different tasks so as to preserve more significant information. 

\subsubsection{Analysis of Time Complexity} 
Now, the computational complexity of the null model-based augmentation is analyzed. Set $n$, $m$, $\alpha$, and $T$ as the number of nodes, the number of edges, the cost coefficient of augmentation, and the approximate augmentation iteration times, respectively, in the original graph. It is easy to verify that the time complexities of 0$k$, 1$k$, 2$k$, LNA, and ADA-(C,  \textsc{Bc}, \textsc{Cc}, \textsc{Ec}) are $\mathcal{O}(n^2)$, $\mathcal{O}(\alpha*m)$, $\mathcal{O}(\alpha*n)$, $\mathcal{O}(m)$, and $\mathcal{O}(T*\alpha*m)$. 

Furthermore, we report the average augmentation running time for each graph based on different augmentation strategies on different datasets. As shown in Table~\ref{tab:runtime}, compared to the training time of the classifier, the 1$k$, 2$k$, and LNA augmentation methods all take less than 0.1 seconds but achieve relatively high improvements (see the classification results in Table~\ref{Tab:NullModel_compare}), while the time consumptions of ADA-C, ADA-\textsc{Bc}, ADA-\textsc{Cc}, and ADA-\textsc{Ec} are comparatively high. The most critical operation is that the corresponding feature value must be recalculated in the augmentation of each edge. While for networks with high dimensionality and lots of edges, the time consumption on feature value calculation could also be large. 

\section{Conclusion}\label{sec:conclusion}
In this paper, we combine the null model with data augmentation to propose several data augmentation methods, which effectively improve the accuracy of graph classification. We conduct experiments to verify the effectiveness of our methods and analyze the experimental results to demonstrated the new findings. We compare five benchmark networks and our results show that the application of data augmentation based on the null model can indeed significantly improve the accuracy of graph classification. We conclude that the null model can be applied to complex networks analysis, and it has great potential in the field of graph mining algorithms design.  
Furthermore, based on the experiments results, we find that the null models have the ability to maintain features consistently with better performance than other methods. These findings also indicate that network features are very important in graph tasks and can provide inspiration for graph data mining research.

In the future, we will study more important features of graph data in graph classification and explore the augmentation methods with such important features to achieve more efficient augmentation. Moreover, we will combine more excellent graph data mining methods with null models in other application scenarios.

\bibliographystyle{IEEEtran}
\bibliography{ref}

\end{document}